\pdfoutput=1
\documentclass[floatfix,a4paper,aps,prd,twocolumn,preprintnumbers]{revtex4}
\usepackage{epsfig}
\usepackage{amsmath}
\usepackage{amssymb}
\usepackage{amsfonts}
\usepackage{graphicx}
\usepackage{graphics,subfigure}
\usepackage{float}
\usepackage{booktabs}
\usepackage{color}
\usepackage{geometry}
\usepackage{rotating}
\usepackage{units}
\usepackage{slashed}
\usepackage{xspace}
\usepackage{ulem}
\usepackage[T1]{fontenc}
\usepackage[utf8]{inputenc}
\usepackage{hyperref}

\hypersetup{
  pdftitle={Probing Minimal Flavor Violation with Long-Lived Stops and
    Light Gravitinos at Hadron Colliders},
  pdfauthor={Jong Soo Kim, Henning Sedello},
}

\newcommand{\bea}{\begin{eqnarray}}
  \newcommand{\eea}{\end{eqnarray}}
\newcommand{\beq}{\begin{equation}}
  \newcommand{\eeq}{\end{equation}}
\newcommand{\beqa}{\begin{eqnarray}}
  \newcommand{\eeqa}{\end{eqnarray}}
\newcommand{\bit}{\begin{itemize}}
  \newcommand{\eit}{\end{itemize}}

\newcommand{\grav}{\ensuremath{\tilde{G}}}
\newcommand{\mgrav}{\ensuremath{m_{3/2}}}
\newcommand{\mpl}{\ensuremath{m_{\rm{Pl}}}}
\newcommand{\st}{\ensuremath{\tilde{t}_1}}

\newcommand{\neutr}{\ensuremath{\tilde{\chi}_1^0}}

\newcommand{\charg}{\ensuremath{\tilde{\chi}_1}}
\newcommand{\gev}[1]{\unit[#1]{GeV}}
\newcommand{\ev}[1]{\unit[#1]{eV}}
\newcommand{\kkev}[1]{\unit[#1]{keV}}
\newcommand{\br}[2]{\mbox{\ensuremath{\mathcal{B}(#1\rightarrow#2)}}}
\newcommand{\dr}[2]{\ensuremath{\Gamma(#1\rightarrow#2)}}

\newcommand{\tev}[1]{\unit[#1]{TeV}}
\newcommand{\kev}[1]{\unit[#1]{keV}}
\newcommand{\ps}[1]{\unit[#1]{ps}}
\newcommand{\ns}[1]{\unit[#1]{ns}}
\newcommand{\fs}[1]{\unit[#1]{fs}}
\newcommand{\met}{\ensuremath{\slashed{E}_T}}

\newcommand{\ifb}[1]{\ensuremath{\unit[#1]{fb^{-1}}}}
\newcommand{\ipb}[1]{\ensuremath{\unit[#1]{pb^{-1}}}}
\newcommand{\order}[1]{\mathcal{O}(#1)}
\newcommand{\dzero}{D$\slashed{\text{0}}$\xspace}
\newcommand{\prog}[1]{\texttt{#1}}

\begin{document}

\title{Probing Minimal Flavor Violation with Long--Lived Stops and
  Light Gravitinos at Hadron Colliders}

\author{J. S. Kim} \affiliation{Institut f\"ur Physik, Technische
  Universit\"at Dortmund, D-44221 Dortmund, Germany\\ ARC Centre of
  Excellence for Particle Physics at the Terascale, School of
  Chemistry and Physics, University of Adelaide, Adelaide, Australia}

\author{H. Sedello}
\affiliation{Institut f\"ur Physik, Technische Universit\"at Dortmund,
  D-44221 Dortmund, Germany}

\begin{abstract}
  In the framework of minimal flavor violation (MFV), we discuss the
  decay properties of a supersymmetric scalar top (stop) in the
  presence of a light gravitino. Given a small mass difference between
  the lighter stop and lightest neutralino and an otherwise
  sufficiently decoupled spectrum, the stop may be long--lived and
  thus can provide support to MFV at hadron colliders. For a
  bino--like lightest neutralino, we apply bounds from searches in the
  $\gamma\gamma\met$ channel (ATLAS with \ifb{1} and \dzero{} with
  \ifb{6.3}) and give a \ifb{5} projection for the ATLAS search.
\end{abstract}

\preprint{DO-TH~11/28}
\preprint{ADP-11-42/T764}

\maketitle

\section{Introduction}
\label{sec:intro}

Supersymmetry (SUSY)~\cite{Ferrara:1974ac} is an attractive extension
of the standard model (SM). However the simplest version of a
supersymmetric SM, the minimal supersymmetric standard model (MSSM)
\cite{Haber:1997if}, does not predict a specific flavor structure; all
superrenormalizable soft supersymmetry breaking terms allowed by gauge
and Lorentz symmetry as well as $R$ parity are present in its Lagrange
density~\cite{Haber:1997if}.  However, it is clear that a
supersymmetric extension of the standard model must have a
non--generic flavor structure to be compatible with experimental
results \cite{Gabbiani:1996hi,Amsler:2008zzb}.

The way the standard model flavor structure is extended to the MSSM is
not unique, yet a widely discussed flavor scheme is minimal flavor
violation~\cite{D'Ambrosio:2002ex} (MFV). In MFV the standard model
Yukawa couplings are promoted to spurion fields transforming under the
SM flavor group to restore the SM's flavor symmetry. If all additional
flavor structure of a new physics model can be understood as higher
dimensional flavor invariant operators including these spurions and
the model's fields, the model is called MFV.

As the LHC is running and eventually will find supersymmetry, it will
be challenging to investigate the flavor structure at a hadron
collider due to the detectors' limited flavor identification abilities
and the complexity of the recorded events. In~\cite{Hiller:2008wp} it
was pointed out that the third generation's squarks decouple from the
first two generations in MFV. As a result, a light stop can be
long--lived decaying through the flavor changing neutral current
channel~\cite{Hikasa:1987db,Muhlleitner:2011ww}
\begin{equation}
  \st\rightarrow c\neutr,
  \label{eq:mfv-decay}
\end{equation}
if all flavor diagonal channels are kinematically closed. ($\st$
denotes the light stop, $\neutr$ the lightest neutralino and $c$ a
charm quark.) An observation of long--living light stops thus would
hint in the direction of MFV.

In MFV, the coupling $Y$ between $\st$, $c$ and $\neutr$ is
\begin{equation}
  Y\propto\lambda_b^2 V_{cb}V_{tb}^*,
  \label{eq:defY}
\end{equation}
where $\lambda_b$ and $V_{ij}$ denotes the bottom Yukawa coupling and
elements of the Cabibbo Kobayashi Maskawa (CKM) matrix
respectively. The precise value of $Y$ depends on the stop left--right
composition, the neutralino decomposition, and on a numeric factor
stemming from the MFV expansion; see Ref.~\cite{Hiller:2008wp} for
details.

In~\cite{Hiller:2009ii} it is shown that the average transverse impact
parameters for the stop decay products can be expected to be
$\order{\unit[1800]{\mu m}}$ for stop lifetimes of the order of ten ps
in the production channel $pp\rightarrow\bar t\bar t
\st\st$~\cite{Kraml:2005kb} .  When both top quarks in this channel
decay leptonically, the pair of same signed leptons in the final state
allows to separate the signal process from its SM background; however,
the small leptonic branching ratio of top quarks suppresses this
process so that, after applying all kinematic cuts, only few events
are left in this channel.  Ref.~\cite{Carena:2008mj} proposes an
alternative collider signature assuming stop pair production in
association with one hard jet. Demanding a minimum transverse momentum
of 1 TeV for the additional jet, the whole parameter region consistent
with electroweak baryogenesis can be
probed. Ref.~\cite{Bornhauser:2010mw} considers an analogous process,
stop pair production in association with two b--jets. However, these
two studies do not consider stops in the MFV framework.

If we consider local SUSY instead of a global implementation of SUSY,
we can have distinct collider signal signatures with little SM
background: In local SUSY, a massive gravitino emerges in the
supersymmetric mass spectrum~\cite{Nilles:1983ge}. Its interactions
with other particles are severely suppressed by the reduced Planck
mass
\begin{equation}
  \mpl=(8\pi G_N)^{-\frac{1}{2}}=2.4\times\gev{10^{18}},
\end{equation}
where $G_N$ is Newton's constant. Depending on the exact breaking
mechanism in the hidden sector, the gravitino can be very light.  A
light gravitino  interacts through its goldstino components
with couplings proportional to
\begin{equation}
  (\mgrav \mpl)^{-1},
\end{equation}
where $\mgrav$ is the gravitino mass.

In models of gauge mediation
\cite{Dine:1981za,Dine:1981gu,Dine:1993yw,Meade:2008wd,Buican:2008ws},
$\mgrav$ is generally much smaller than the sparticle mass scale;
thus, the gravitino is the lightest supersymmetric particle (LSP). Its
goldstino interactions are enhanced and can be of the electroweak
order.

Consequently, the lightest neutralino decays via
\begin{equation}
  \neutr\rightarrow X\grav,
  \label{eq:neutr-decay}
\end{equation}
where $X$ denotes a photon, $Z$, or a
Higgs boson; $\grav$ denotes a gravitino~\cite{Ambrosanio:1996jn}. If
$X$ is a photon ($\gamma$), this decay leads to very clear collider
signatures with high $p_T$ isolated photons plus missing transverse
energy ($\met$) stemming from gravitinos leaving the detector unseen.

Several studies with light gravitinos at hadron colliders were
performed in the
past. Ref.~\cite{Shirai:2009kn,Baer:1998ve,Ambrosanio:1996jn} consider
the diphoton plus $\met$ channel at hadron colliders. In
Ref.~\cite{Hamaguchi:2006vu,Ellis:2006vu} the authors examine a stau
NLSP and a gravitino LSP. A sneutrino NLSP and a gravitino LSP
scenario is investigated in
Ref.~\cite{Katz:2009qx,Santoso:2009qa}. Ref.~\cite{Meade:2009qv}
considers the discovery potential of a neutralino NLSP and a gravitino
LSP at the Tevatron, where they consider a general decomposition of
the lightest neutralino. In Ref.~\cite{Ambrosanio:1996jn} the authors
consider a light stop NNLSP and a light neutralino NLSP and a
gravitino LSP. Ref.~\cite{Kats:2011it} investigate a stop NLSP and a
gravitino LSP scenario for the Tevatron as well as the LHC. A chargino
NLSP and a gravitino LSP is considered in
Ref.~\cite{Kribs:2008hq}. Depending on the size of the gravitino mass,
non--pointing photons can be measured. The discovery potential of
sparticle decays with a finite decay length are investigated in
\cite{Hamaguchi:2007ji,Feng:2010ij,Meade:2010ji}. A recent
experimental search for sparticles with finite decay lengths is
published in~\cite{Aad:2011zb}.

In this paper we investigate the parameter region where the decay in
Eq.~\eqref{eq:mfv-decay} is dominant in the context of a light
gravitino, and how the stop masses are constrained in this framework
by recent collider searches, assuming that $\st$, $\neutr$, and
$\grav$ are the only light supersymmetric particles. We discuss the
stop and $\neutr$ decay patterns in section~\ref{sec:patterns}. In
section~\ref{sec:collider} we apply collider bounds in the
$\gamma\gamma\met$ channel~\cite{Abazov:2010us,arXiv:1111.4116}.  The
cuts adapted from the experimental studies are discussed in the two
appendices.

\section{Decay patterns}
\label{sec:patterns}
\subsection{Light stop decays}
\label{sec:stop}

Here we discuss possible decay patterns of light stops and their
implications on the parameter space. We start with decays of the light
stop via Yukawa and gauge couplings and then discuss direct stop
decays into a gravitino.

The lighter stop mass eigenstate $\st$ is the lightest squark state in
many supersymmetry breaking scenarios. On one hand, the large top
Yukawa coupling can induce a sizable left-right mixing in the stop
sector leading to light $\st$ masses. On the other hand, if the soft
squark mass terms are unified at a high scale, the stop mass terms are
prominently reduced by the top Yukawa coupling in the running of the
renormalization group equations to the electroweak
scale~\cite{Ibanez:1984vq}. In this case, $\st$ is mostly a $SU(2)$
singlet state since its mass term does not receive any contributions
from $SU(2)$ gaugino loops.  In addition, right--handed stop loop
contributions to the rho parameter are sufficiently
suppressed~\cite{Drees:1990dx}.

Surprisingly light masses of $\order{\gev{100}}$ are still consistent
with experimental searches for stops decaying to $c\neutr$ at the
Tevatron if the mass splitting
\begin{equation}
  \Delta m = m_{\st}-m_{\neutr}
  \label{eq:massSplitting}
\end{equation}
is smaller than~$\approx\gev{30}$~\cite{CDFexotic}.

Since we want the flavor changing decay in Eq.~\eqref{eq:mfv-decay} to
be the dominant decay and the stop to be long--lived, we must ensure
that the potentially dominant decays $\st\rightarrow t \neutr$,
$\st\rightarrow b\charg^\pm$, and $\st\rightarrow b\neutr W$ are
kinematically closed or sufficiently suppressed,
\begin{equation}
  m_{\st}<m_b+m_{\neutr}+ m_{W},\quad m_{\st}<m_b+m_{\charg^\pm}.
\end{equation}

Depending on the chargino and slepton masses~\mbox{($>m_{\st}$)}, the
four--body decay $\st\rightarrow b \ell \nu\neutr$ may be dominant if
$\Delta m$ exceeds a few \gev{10}. For small stop neutralino mass
splittings, the tree--level four--body decay is strongly phase space
suppressed and Eq.~\eqref{eq:mfv-decay} remains the dominant decay
mode~\cite{Hiller:2008wp}.

The MFV decay width can be written as
\begin{equation}
  \label{eq:sigdecay}
  \Gamma(\st\rightarrow c\neutr) = \frac{m_{\st}
    Y^2}{4\pi}\left(\frac{\Delta
      m}{m_{\st}}\right)^2
\end{equation}
in the limit of $\Delta m \ll m_{\st}$~\cite{Hiller:2008wp}. Thus, if
this decay is dominant, the stop lifetime is governed by $\Delta m
Y$. The kinetic energy of the hadronic stop decay remnants depends on
the size of the mass splitting $\Delta m$. In a study with same-sign
leptons~\cite{Hiller:2009ii}, a major reduction of event numbers due
to a minimal $p_T$ cut on these low energy decay products has been
found.

In the presence of a light gravitino, restrictions on $\Delta m$ and
$Y$ are not sufficient to guarantee the dominance of the FCNC (flavor
changing neutral current) decay mode Eq.~\eqref{eq:mfv-decay}.  If
$\Delta m Y$ is too small, the two--body decays of the stop into the
gravitino,
\begin{equation} \st\rightarrow\grav c,\quad\st\rightarrow \grav t,
  \label{eq:2bGrav}
\end{equation}
and---if the resonant decay to tops is kinematically closed---the
three--body decay mode of the stop,
\begin{equation}
  \st\rightarrow \grav b W,
  \label{eq:3bGrav}
\end{equation}
can have a significant contribution to the full stop decay
width. These decays can invalidate our assumption, that the stop
dominantly decays via a FCNC process with a finite impact parameter.

For the flavor diagonal two and three--body decay channels, the decay
rates are~\cite{Ambrosanio:1996jn,Sarid:1999zx}
\begin{subequations}
  \label{eq:gravdecay}
  \begin{align}
    \label{eq:grav2bdecay}
    \Gamma(\st\rightarrow t\grav) & =
    \frac{1}{48\pi}\frac{m_{\st}^5}{\mpl^2\mgrav^2}
    \left(1-x_t^2\right)^4\\
    \label{eq:grav3bdecay}
    \Gamma(\st\rightarrow W^+b\grav) & =
    \frac{V_{tb}^2\alpha_{em}}{384\pi^2\sin^2{\theta_W}}
    \frac{m_{\st}^5}{\mpl^2\mgrav^2}\nonumber\\ 
    \cdot\left[|c_L|^2\right.I\left(\right.&\!\!\!\left.\left.x_W^2,x_t^2\right)
      +|c_R|^2J\left(x_W^2,x_t^2\right)\right],
  \end{align}
\end{subequations}
where the gravitino mass is neglected in the phase space
integrals. $\alpha_{em}$, and $\theta_W$ denote the fine-structure
constant and the Weinberg angle, respectively. Further $x_W=m_W/m_{\st}$
and $x_t=m_t/m_{\st}$ with the top mass $m_t$. $c_L$ and $c_R$
parametrize the $\tilde{t}_L$ and $\tilde{t}_R$ contribution to $\st$;
due to the 3rd generation's decoupling in MFV the other squarks'
admixture is small, \textit{i.e.}  $|c_L|^2+|c_R|^2\approx 1$. The
functions $I(x_W^2,x_t^2)$ and $J(x_W^2,x_t^2)$ are phase space
integrals and can be found in~\cite{Sarid:1999zx}.  Note that
Eq.~\eqref{eq:grav3bdecay} does not comprise the finite top width and
diverges at the top mass threshold. We use this formula for
$m_{\st}<m_t$ only. As the three--body decay proceeds trough a virtual
top quark, its rate is largest for a right--handed stop, because the
chirality flipping top mass dominates the propagator.

For a bino--like $\neutr$, the flavor structure of the $\st-\grav-c$
coupling stemming from the MFV expansion is the same as in the
$\st-\neutr-c$ coupling, thus the decay rate for $\st\rightarrow \grav
c$ can be written as~\cite{Hiller:2009ii}
\begin{equation}
  \label{eq:cgravdecay}
  \dr{\st}{\grav c}=
  \frac{Y^{\prime^2}}{48\pi}\frac{m_{\st}^5}{m_{\rm{Pl}}^2\mgrav^2},
\end{equation}
where $Y'$ and $Y$ are related by a factor dependent on the stop
composition as $Y$ comprises the hypercharges of the left- and
right--handed stop fields. The factor is
\begin{equation}
  \left|\frac{Y'}{Y}\right|=\frac{1}{\sqrt{2}g'Y_Q}\approx
  \begin{cases}
    3&\text{(right--handed $\st$)}\\
    12&\text{(left--handed $\st$)}
  \end{cases}
\end{equation}
with the SM ${U}(1)$ coupling $g'$ and the left--handed
(right--handed) stop hypercharge
$Y_Q=\frac{1}{6}\left(\frac{2}{3}\right)$.

\begin{figure}[ht]
  \input{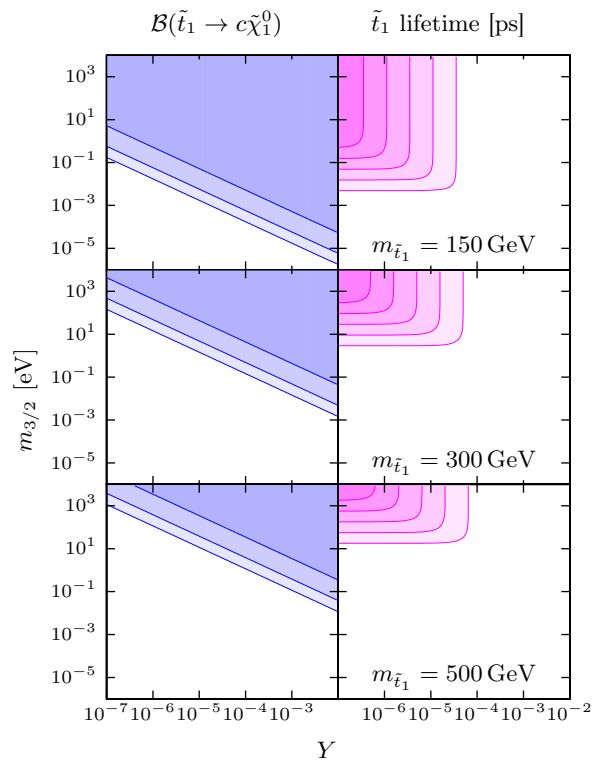}
  \caption{MFV light stop branching ratios (left, blue) and stop
    lifetimes (right, pink), respectively for $m_{\st}=\gev{150}$,
    $\gev{300}$ and $\gev{500}$ (from top to bottom), in dependence on
    the gravitino mass $m_{\grav}$ and the MFV coupling $Y$. In the
    branching ratio plots, the shaded areas correspond to
    $\br{\st}{c\neutr}\ge0.01$, $0.1$, $0.9$ (from light to dark hue),
    whereas, in the lifetime plots, the shaded areas correspond to
    $\tau_{\st}\ge0.01$, $0.1$, $1$, $10$, $\ps{100}$ (from light to
    dark hue). We chose $\Delta m=\gev{10}$.}
  \label{fig:brs}
\end{figure}

We show the branching ratio $\br{\st}{c\neutr}$ and the stop lifetime
in the $m_{\grav}$--$Y$ plane for three different masses of a
right--handed stop in Fig.~\ref{fig:brs} using the decay
rates~\eqref{eq:sigdecay}, \eqref{eq:gravdecay},
and~\eqref{eq:cgravdecay}.  To generate the plots, we keep $\Delta m$
fixed at $\gev{10}$ and use $\sin^2\theta_W=0.23$,
$\alpha_{em}=1/128$, and $m_t=\gev{173}$. The plot for left--handed
stops does not differ significantly from the one shown.

Due to the $m_{\st}^5$ dependence of the decay widths in
Eqs~\eqref{eq:2bGrav},~\eqref{eq:3bGrav} and the weaker $m_{\st}^{-1}$
dependence of $\Gamma(\st\rightarrow c\neutr)$, the minimal gravitino
mass necessary to account for a sizable \br{\st}{c\neutr} increases
with larger stop masses. The $m_{\st}^{-1}$ dependence of
$\Gamma(\st\rightarrow c\neutr)$ also causes the smallness of the
shifts of the lifetime regions in Fig.~\ref{fig:brs} to larger $Y$
values when the stop mass is increased.

As it is clearly visible from Fig.~\ref{fig:brs}, very small values of
$Y\lesssim\order{10^{-5}}$ and at least gravitino masses of
$\order{\kev{0.1-1}}$ are required in addition to the mass hierarchy
\begin{equation}
  \mgrav\ll m_{\neutr}\leq m_{\st}\leq m_{\charg^\pm}
  \label{eq:massOrder2}
\end{equation}
for the stop to be long--lived and to decay dominantly to $c\neutr$.

As the charmed hadron produced in the decay has a macroscopic lifetime
of $\order{\ps{1}}$ itself, however, a macroscopic stop lifetime might
turn out to be accessible experimentally only if it exceeds this
timescale.

\subsection{NLSP neutralino composition and decays}
\label{sec:neutralino}
Neutralinos are mass eigenstates of the $U(1)$ gauge fermion (bino),
the neutral $SU(2)$ gauge fermion (wino), and the neutral up-- and
down--type Higgs fermion (higgsino). The fields' individual
contributions to $\neutr$ as well as the neutralino mass spectrum
depends on the bino mass $M_1$, the wino mass $M_2$, the Higgs mixing
parameter $\mu$ and the ratio between the up--type and down--type
Higgs vacuum expectation value (VEV) $\tan\beta$. If $\neutr$ is the
NLSP and $\grav$ the LSP, $\neutr$ decays via $\neutr\rightarrow X
\grav$, where $X$ is either the photon, the Z boson, or a neutral
Higgs boson. Branching ratios into the various decay channels are
fixed by phase space suppression factors and the decomposition of the
lightest neutralino. General formulae for the decay widths are given
in Refs~\cite{Ambrosanio:1996jn,Meade:2009qv}.

% what about ...
% ...wino
In the previous subsection, we argued that the mass of the light
chargino $\charg^\pm$, a mass eigenstate of charged winos and
higgsinos, has to be larger than the light stop mass in order to
suppress the flavor diagonal stop decay to $\charg^\pm b$. This
requirement cannot be satisfied if the $\neutr$ is wino--like,
\textit{i.e.} if $M_2\ll M_1,|\mu|$, as in this case both the $\neutr$
and the $\charg^\pm$ mass are~$\approx M_2$. The mass splitting
between wino--like $\charg^\pm$ and $\neutr$ is of the order of
$\frac{m_Z^5}{\mu^4}$~\cite{Martin:1993ft}, given $M_1 \ll |\mu|$, and
thus is extremely suppressed for $|\mu|\gtrsim\text{few }\gev{100}$.

% ...higgsino
Similarly, if $\neutr$ is higgsino--like, $\neutr$ and $\charg^\pm$
have masses of the same order of magnitude given by $\mu$.  The mass
splitting $\Delta m_{\charg^\pm\neutr}$ is of the order of
$\frac{m_Z^2}{M_2}$~\cite{Martin:1993ft} for $|\mu|\ll M_1, M_2$.

% ...bino
If $\neutr$ is bino--like, its mass is $M_1$ approximately, while the
$\charg^\pm$ mass is given by $|\mu|$ or $M_2$. As the mass gap
depends on two different supersymmetric mass parameters, it can be
sizable depending on the details of high scale physics.

% why bino: 1) stable decay chain w.r.t. the mass scale. 2) have photons
While in a mass region close to the $Z$ mass, also a higgsino--like
$\neutr$ may respect the anticipated mass hierarchy in
Eq.~\eqref{eq:massOrder2}, we focus on a bino--like $\neutr$ in
discussing experimental bounds as 1)~in the bino case the mass
hierarchy can exist over a large stop mass scale and 2)~binos have a
large branching fraction to photons. For $m_{\neutr}>m_{Z}$ and
negligible phase space suppression, the branching ratio is
$\br{\neutr}{\gamma\grav}\approx\cos^2\theta_W$. This value is obvious
as $\grav$ is a gauge singlet and $\gamma$ a mixed state of the
hypercharge gauge boson and the neutral $SU(2)$ gauge boson where the
mixture is parametrized by the Weinberg angle.  Including the phase space
suppression from $m_{Z}$ and assuming that the higgsino sector is
decoupled, the bino decay rates are~\cite{Ambrosanio:1996jn,hep-ph/9609434}
% - bino decay 
\begin{subequations}
  \label{eq:neutrdecay}
  \begin{align}
    \Gamma(\neutr\rightarrow \gamma\grav) &=
    \cos^2{\theta_W}\frac{m_{\neutr}^5}{48\pi\mgrav^2\mpl^2}\\ \Gamma(\neutr\rightarrow
    Z\grav) &=\nonumber\\
    \sin^2&\theta_W\frac{m_{\neutr}^5}{48\pi\mgrav^2\mpl^2}\left(1-\frac{m_{Z}^2}{m_{\neutr}^2}\right)^4.
  \end{align}
\end{subequations}

\begin{figure}[ht]
  \input{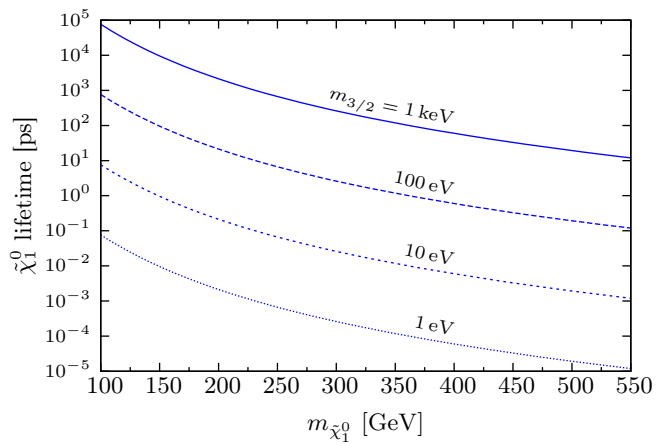}
  \caption{Lightest neutralino lifetime as a function of the
    neutralino and gravitino mass.}
  \label{fig:neutr-ltime}
\end{figure}

For reference we show the bino--neutralino lifetime as a function of
the lightest neutralino mass for gravitino masses 1, 10, 100, 1000 eV in
Fig.~\ref{fig:neutr-ltime}.

\section{Collider bounds}\label{sec:collider}

When $\st$, $\neutr$, and $\grav$ are the only light
supersymmetric particles, stops are dominantly produced in pairs, both
at $\bar pp$ and $pp$ colliders,

\begin{equation}
  \bar pp\rightarrow\st\st^*,\qquad pp\rightarrow\st\st^*.
\end{equation}

The production cross sections are given in Fig.~\ref{fig:xsec} as a
function of the stop mass for the LHC at \tev{7} as well as for
Tevatron and are calculated with
\prog{Prospino}~\cite{Beenakker:1997ut} using the built-in
CTEQ6.6M~\cite{Nadolsky:2008zw} parton distribution functions (PDFs).
We also show the next--to--leading order uncertainty by varying the
factorization scale ($\mu_F$) and the renormalization scale ($\mu_R$)
between $\frac{1}{2}m_{\st}$ and $2m_{\st}$ while keeping $\mu_R$
equal to $\mu_F$.

Given a bino--like $\neutr$, the final state
signatures of a decay chain via Eq.~\eqref{eq:mfv-decay} and
Eq.~\eqref{eq:neutr-decay} are
\begin{equation}
  \gamma\gamma c\bar c\grav\grav,\quad
  \gamma Z c\bar c\grav\grav,\quad
  Z Z c\bar c\grav\grav.
\end{equation}
In general, with a small mass gap $\Delta m$, the charm jets are too
soft to be useful for event selection.  Thus constrains on our
parameter space can stem from searches for an excess in the
$\gamma\gamma\met$, $\gamma Z\met$ and $ZZ\met$ channels. As binos
dominantly decay to $\gamma\grav$, the SM background is negligible for
energetic photons, and large $\met$ and high $p_T$ photons are
efficiently identified in multi\-purpose detectors, we concentrate on
the $\gamma\gamma\met$ channel in this work. Several experimental
searches for the diphoton and $\met$ channel have been
published~\cite{Aad:2010qr, Aaltonen:2009tp,Chatrchyan:2011wc,
  Abazov:2010us, arXiv:1111.4116}. So far, no excess above the SM
expectation has been found.

In the following, we present exclusion limits in the stop--gravitino
mass plane derived from the latest \mbox{ATLAS} search in the
$\gamma\gamma\met$ channel for a luminosity ($\mathcal L$) of
$\ifb{1.07}$~\cite{arXiv:1111.4116}. We derive also bounds from the
\dzero{} search with $\mathcal L=\ifb{6.3}$~\cite{Abazov:2010us}, and
give a $\mathcal L=\ifb{5}$ projection for the \mbox{ATLAS} bound.

The dominant SM background with $\met$ originating from the hard
process stems from $W+\gamma$, $W+{\rm jets}$, and $W/Z\gamma(e)
\gamma(e)$ production. Here, electrons/jets are misidentified as
photons. SM background with $\met$ from mismeasurements emerges from
multijet production and direct photon production.

A supersymmetric background can only arise from $\neutr\neutr$ pair
production as all other colored sparticles, sleptons, heavier
neutralinos and charginos are assumed to be heavy and thus will have a
negligible contribution.  However, also the $\neutr\neutr $ production
cross section is severely suppressed. In the limiting case of
vanishing higgsino admixture to $\neutr$, the cross section vanishes
even at $\order{\alpha_{\text{EW}}^2\alpha_s}$. Consequently we do not
take $\neutr\neutr$ pair production into account in the following.

\begin{figure}
  \input{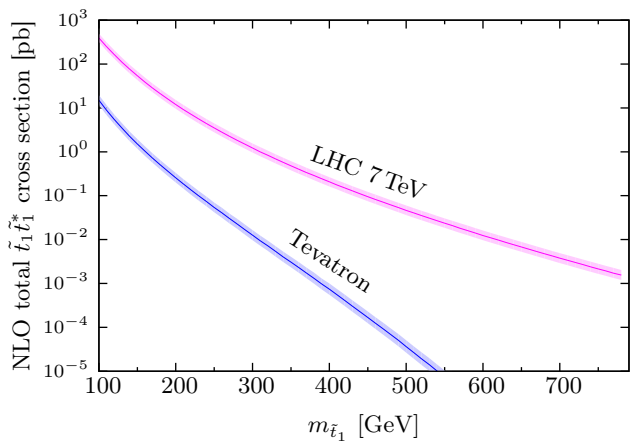}
  \caption{NLO total cross section for $\st\st^*$ production at the
    Tevatron and at the LHC with $\sqrt{s}=\tev{7}$. The colored bands correspond to scaling the unified scale
    $\mu_F=\mu_R=m_{\st}$ by $\frac{1}{2}$ and $2$.}
  \label{fig:xsec}
\end{figure}

\subsection{Calculation of exclusion limits}

To constrain the stop mass, the gravitino mass, and the MFV coupling
$Y$, we calculate $\sigma_{\st\st^*}$, the total cross section for
$\st\st^*$ production, in a grid of the light stop mass $m_{\st}$
using \prog{Prospino}. Note here that the light stop mass is the
dominant SUSY parameter in the cross section~\cite{Beenakker:1997ut},
both for $p\bar p$ and $pp$ initial states.

For each stop mass in the grid, we generate 100$\,$000 $\st\st^*$ pair
events with \prog{pythia~6.4.25}~\cite{Sjostrand:2006za} using the
CTEQ6.6M~\cite{Nadolsky:2008zw} parton distribution functions. With the
hadron level events we simulate the efficiency/acceptance for the
$\gamma\gamma c\bar c\grav\grav$ final state in the \mbox{ATLAS} and
\dzero{} analyses employing a slightly modified version of
\prog{Delphes~1.9}~\cite{Ovyn:2009tx}\footnote{We modify \prog{Delphes}
  slightly to simulate a \dzero-like calorimeter with $>40$ segments
  in $\eta$ direction and flag gravitinos as undetectable particles.}.
The photon energy, and therefore our signal's detection efficiency,
depends on the mass splitting $\Delta m$. As in the previous sections,
we fix $\Delta m =\gev{10}$. In appendices~\ref{sec:dzero}
and~\ref{sec:atlas}, we give details on the cuts adopted from the
experimental studies and on further simulation parameters.  As a
result of this simulation step, we obtain efficiencies $\epsilon_n$ in
bins of $\met$ and can calculate a signal cross section in bin $n$
from
\begin{equation}
  \sigma^{sig}_{n}=\epsilon_n\br{\st}{\neutr c}^2\br{\neutr}{\grav
    \gamma}^2\sigma_{\st\st^*}.
\end{equation}

Using Eqs~\eqref{eq:neutrdecay} to calculate
$\br{\neutr}{\grav\gamma}$, we finally employ the $\mathrm {CL}_s$
method~\cite{Junk:1999kv,599622}\footnote{We use the implementation in
  the \prog{TLimit} class of \prog{ROOT 5.28.00b}\cite{Brun:1997pa}.}
to calculate the 95\% exclusion limits for $\br{\st}{\neutr c}$. Those
are depicted in Fig.~\ref{fig:br-excl}, where we use the measurements
plus background predictions of the experimental studies as enlisted in
Tab.~\ref{tab:data}.  When calculating the exclusion limit, we treat
the errors on the luminosity and the background as Gaussian nuisance
parameters, see Tab.~\ref{tab:data}, but do not take into account
theory uncertainties stemming from scale variations and the choice of
PDF sets.

The projection for the ATLAS study with a luminosity of $\ifb{5}$ is
calculated using the prescription of~\cite{Conway:2000ju}.

\begin{table}[h]
  \begin{flushleft}
    \begin{tabular}{l|c|r|r}
      & $\met$ Bin [\unit{GeV}]& \parbox{4.1em}{Observed events}
      & \parbox{4.5em}{SM bgd events}\\
      \hline
      \dzero{}~\cite{Abazov:2010us}&$35-50$& $18$&$11.9\pm2.0$\\
      $\ifb{(6.3\pm0.4)}$&$50-75$&$3$&$5.0\pm0.9$\\
      &$>75$&$1$&$1.9\pm0.4$\\
      \hline
      ATLAS~\cite{arXiv:1111.4116}&$>125$&$5$&$4.1\pm0.6$\\
      $\ifb{(1.07\pm0.04)}$&&&\\
    \end{tabular}
  \end{flushleft}
  \caption{ATLAS and \dzero measurements and background (bgd) predictions.}
  \label{tab:data}
\end{table}

\begin{figure}
  \input{br-bounds_t}
  \caption{Maximal branching ratio for $\st\rightarrow\neutr c$ in
    dependence of the stop mass given a bino-like $\neutr$ and a fixed
    stop neutralino mass difference of $\Delta m=\gev{10}$.  Blue
    curve and blue shaded area: \dzero
    study~\cite{Abazov:2010us}. Pink curves: ATLAS
    search~\cite{arXiv:1111.4116}, where the solid curve corresponds
    to the measurement, and the dashed line corresponds to a $\mathcal
    L=\ifb{5}$ projection. The areas above the curves are excluded at
    95\% CL within our simplified analysis.}
  \label{fig:br-excl}
\end{figure}

\subsection{Numerical analysis and discussion}

As can be seen in Fig.~\ref{fig:br-excl}, the ATLAS search (solid pink
curve) gives a bound on $\br{\st}{\neutr c}$ for stop masses up to
$\gev{560}$.  For a luminosity of $\ifb{5}$, this mass is projected to
$\gev{660}$.  For larger masses, a dominant tree level FCNC decay
$\st\rightarrow\neutr c$ is in agreement with the measurement. As can
be seen from Fig.~\ref{fig:br-excl}, the \ifb{1.07} ATLAS data already
excludes a larger parameter region than the \ifb{6.3} \dzero{}
data~\footnote{We also performed the calculation for the \ipb{36} CMS
  $\gamma\gamma\met$ results~\cite{Chatrchyan:2011wc}. These impose a
  weaker bound on $\br{\st}{c\neutr}$ than the \ifb{6.3} \dzero{}
  data~\cite{Abazov:2010us}.}.

For each stop mass, the bound in Fig.~\ref{fig:br-excl} can be mapped
to a bound on the maximal gravitino mass for given values of $Y$ using
Eqs~\eqref{eq:sigdecay},~\eqref{eq:gravdecay},
and~\eqref{eq:cgravdecay}. We plot these bounds for $Y=10^{-7}$,
$10^{-6}$, $10^{-5}$ in Fig.~\ref{fig:excl}. As
Eqs~\eqref{eq:gravdecay} do not provide the correct stop width for
masses in the threshold region close to the top mass, we exclude this
region from the mapping and interpolate our result in the region
$m_t\pm\gev{30}$ ($m_t=\gev{173}$).

The mass difference $\Delta m$ enters the bounds in
Fig.~\ref{fig:excl} through Eq.~\eqref{eq:sigdecay} and through the
hardness of the photon $p_T$ spectrum. For small changes in $\Delta
m$, the latter dependency can be neglected, and the bounds depend on
the product $(Y\Delta m)$ only. Therefore the bounds for other viable
values of $\Delta m$ can be derived from those shown for $\Delta m
=\gev{10}$ by rescaling $Y$.

Obviously, the smaller we choose $Y$ the larger $\mgrav$ can be to
generate a branching ratio below a certain value. The bound on
$\mgrav$ resulting from the mapping in Fig.~\ref{fig:excl} varies over
a wide mass range and thus potentially implies different regimes of
$\st$ and $\neutr$ lifetimes; therefore, we show slopes of fixed
values: For the $\neutr$ lifetime, we show slopes for \fs{1}, \ps{1},
and \ns{1} (black dotted) following approximately $\mgrav\propto
m_{\st}^{5/2}$. For $\st$, we show slopes for $\ps{1/5}$, $\ps{1}$,
and $\ps{5}$ (black solid). At $Y=10^{-5}$, the stop lifetime is below
\ps{1} already induced by $\st\rightarrow \neutr c$ irrespective of
the gravitational decay channels; therefore, only the $\ps{1/5}$ slope
can be drawn here. Similarly, the stop lifetime drops below $\ps{1/5}$
for stop masses smaller than $\approx\gev{230}$ for $Y=10^{-5}$, as
$\Gamma(\st\rightarrow c\neutr)\propto m_{\st}^{-1}$ for fixed $\Delta
m$. For both smaller values of $Y$, the stop lifetime slopes shown
only depend weakly on $Y$ because the corresponding total stop widths
are dominated by the gravitational decay $\st\rightarrow t\grav$
resp.~$\st\rightarrow W^+b\grav$.

\begin{figure}
  % \fbox{
  \input{m32-mass-bounds_t}
  % }
  \caption{Pink curves: The maximal gravitino mass respecting the
    ATLAS measurements~\cite{arXiv:1111.4116} in dependence of the
    light stop's mass. The solid curve corresponds to the measurement
    for $\mathcal L = \ifb{1}$, and the dashed line corresponds to a
    $\mathcal L=\ifb{5}$ projection. Blue curve: $\mathcal L =
    \ifb{6.3}$ \dzero{} search~\cite{Abazov:2010us}. Gravitino masses
    above the curves are excluded at 95\%~CL.  The black dotted curves
    are slopes of fixed $\neutr$ lifetime (\fs{1}, \ps{1}, and \ns{1}
    from bottom to top) for $m_{\neutr}=m_{\st}-\gev{10}$. Along the
    solid black lines, the stop lifetime is \ps{0.2}, \ps{1}, and
    \ps{5}. }
  \label{fig:excl}
\end{figure}

Fig.~\ref{fig:excl} shows that the gravitino mass region promoted
in~\cite{Hiller:2009ii} as the region where the decay
$\st\rightarrow\neutr c$ dominates for stop masses between $100$ and
$\gev{150}$ is now disfavored. More generally,
Fig.~\ref{fig:excl} allows to discuss two regimes of different orders
of $\st$ and $\neutr$ lifetimes:
\begin{itemize}

\item In the stop mass region where $\br{\st}{\neutr c}$ is bounded,
  \textit{i.e.} for $m_{\st}\lesssim\gev{500}$, the gravitino channels
  have a significant contribution to the stop decay. If this
  contribution is dominant, both---the $\neutr$ and the $\st$
  decay---are governed by the same coupling $\sim1/{\mgrav^2}$. As the
  stop decay width is suppressed by phase space ($t\grav$ channel) or
  top propagator ($\grav W^+b$ channel), the stop lifetime is expected
  to be larger than, or at the same order of magnitude as, the
  neutralino lifetime in this region.

\item In the mass region where $\br{\st}{\neutr c}$ is allowed to
  dominate, \textit{i.e.} for $m_{\st}\gtrsim\gev{500}$, the phase
  space suppression of the stop decay width Eq.~\eqref{eq:grav2bdecay}
  is less pronounced; thus, the stop and neutralino gravitational
  partial decay widths are of the same order of
  magnitude. Consequently, if $\st\rightarrow c\neutr$ is the dominant
  stop decay in this mass region, the stop lifetime is significantly
  smaller than the neutralino lifetime.
\end{itemize}

The relation between the stop and neutralino lifetimes ($\tau_{\st}$
and $\tau_{\neutr}$) described above is summarized in
Fig.~\ref{fig:lt-ratios} where we plot the allowed ratio of both
lifetimes within the bounds.  Along the solid black line bounding the
grey area, the contribution of Eq.~\eqref{eq:sigdecay} to the stop
decay width vanishes; thus, this line represents the smallest value
the ratio can have.  The plot is generated for $Y=10^{-6}$; however,
only the leftmost parts of the exclusion areas depend weakly on the
specific value of $Y$. For smaller masses, the bounds are driven by
the phase space dependence of Eqs~\eqref{eq:gravdecay}
and~\eqref{eq:neutrdecay}.

\begin{figure}
  \input{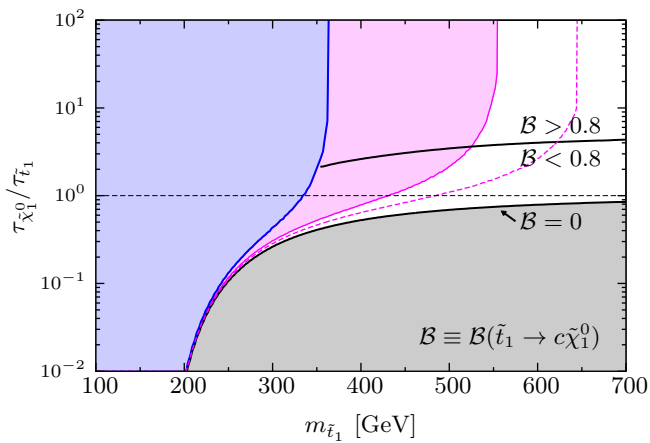}
  \caption{White area: the available ratio of $\neutr$ lifetime over
    $\st$ lifetime for $Y=10^{-6}$, $\Delta m=\gev{10}$. Blue area:
    excluded by the \dzero data~\cite{Abazov:2010us}. Pink area:
    excluded by the \ifb{1.07} ATLAS data~\cite{arXiv:1111.4116}. Pink
    dashed line: \ifb{5} projection for ATLAS. Grey area: invalid.}
  \label{fig:lt-ratios}
\end{figure}

Note that for large $\st$ or $\neutr$ lifetimes, care must be taken in
the interpretation of the bounds in Figs~\ref{fig:excl}
and~\ref{fig:lt-ratios}: The selection criteria for photon candidates
in the ATLAS publication are chosen for prompt
photons~\cite{arXiv:1012.4389,atlas-conf-2010-005}. Also, more
explicitly, the \dzero{} study requires that photon candidates point
to a reconstructed primary vertex. We assumed in our calculation that
all signal photons fulfill these criteria as if they were prompt. In a
more realistic simulation, for increasing neutralino lifetimes, the
signal photons' selection efficiency should decrease. For longitudinal
neutralino decay lengths of $\order{\unit[1000]{mm}}$ ATLAS
simulations show that for several photon selection criteria
\footnote{The ratio of the energy deposits in $3\times7$ and
  $7\times7$ cells ($\eta\times\phi)$ in the electromagnetic
  calorimeter contributing to the photon cluster, and the shower's
  lateral width.} used in~\cite{arXiv:1111.4116} the efficiency drops
from $\order{85\%}$ to $\order{55\%}$~\cite{Aad:2009wy}. Consequently,
for lifetimes $\gtrsim\order{\ns{}}$, we overestimate the number of
photons accepted, and the bounds on $\mgrav$ presented should be
regarded with care in this lifetime region.

%1698

While for a bino--like neutralino, the $\gamma\gamma\met$ channel
offers the highest sensitivity for setting mass limits, it may be
difficult to measure the neutralino lifetime in this channel due to
lack of photon tracks. To construct the photons' impact parameters,
CMS focuses on converted photons in a \ifb{2.1} search in the
$\gamma\gamma\met+\text{jets}$ channel
in~\cite{CMS-PAS-EXO-11-067}. As pointed out in~\cite{Meade:2010ji},
the $\neutr\rightarrow Z\grav$ channel may be used to investigate the
neutralinos' lifetimes, as the $Z$'s decay products allow to
reconstruct the $\neutr$'s trajectory.

\section{Summary}\label{sec:summary}

In SUSY models with MFV, the third generation of squarks decouples
from the other two generations. This decoupling opens the opportunity
to support the MFV hypotheses with the measurement of a macroscopic
stop decay length if the stop decay can only proceed through a
generation--changing channel due to kinematic
constraints~\cite{Hiller:2008wp}. In this work, we investigate the
decay of a stop into a charm and a bino--like lightest neutralino with
a subsequent bino decay to a photon and a light gravitino,
\begin{equation}
  \st\rightarrow\neutr (\rightarrow\gamma\grav) c,
\label{eq:complete_decay}
\end{equation}
given a sufficiently small mass difference between $\st$ and $\neutr$.  

While a macroscopic stop decay length serves as a hint for MFV, the
neutralino's decay to a photon leaves a distinct signature in LHC and
Tevatron detectors offering a good signal isolation.

Assuming that the remainder of the SUSY spectrum is decoupled, we find that
the ATLAS search in the $\gamma\gamma\met$ channel based on a
luminosity of $\ifb{1}$~\cite{arXiv:1111.4116} implies a bound on
$\br{\st}{c\neutr}$ for stop masses up to $\gev{560}$, see
Fig.~\ref{fig:br-excl}. In a $\ifb{5}$ projection, the bound is
raised to $\gev{660}$.

We find that stops lighter than $\sim\gev{400}$ are still compatible
with the $\gamma\gamma\met$ searches; however, in this region, a
significant fraction of the stops decays into gravitinos and quarks of
the third generation. Here the stops are expected to have larger
lifetimes than the lightest neutralinos; though, a macroscopic stop
decay length is governed by the gravitino mass scale $\mgrav$ and is
no hint for a decoupled stop flavor mixing structure.

Stops heavier than $\sim\gev{500}$ can dominantly decay through the
decay chain in Eq.~\eqref{eq:complete_decay} and eventually support
MFV if they are a long--lived. As the lifetime depends on both the
stops's flavor mixing and the gravitino mass, it can serve as a hint
to MFV only if the contribution of the gravitino modes to the total
decay width is negligible. This is assured if the neutralino lifetime
is much larger than the stop lifetime.

We have discussed a split spectrum where the only light particles are
a light gravitino, a light bino, and a light right--handed stop. It
may be possible, however, to separate supersymmetric background
processes in a scenario with additional light sparticles by vetoing on
additional jets and/or isolated leptons.

\begin{acknowledgments}
  We thank Manuel Drees, Sebastian Grab, and Gudrun Hiller for
  discussions in the initial phase of this project and for the
  comments on the manuscript. JSK also thanks the Bethe Center of
  Theoretical Physics and the Physikalisches Institut at the
  University of Bonn for their hospitality.
  This work has been supported in part by the Initiative and Networking Fund
  of the Helmholtz Association, Contract No.~HA-101 (``Physics at the
  Terascale'') and by the ARC Centre of Excellence for Particle
  Physics at the Terascale.
\end{acknowledgments}

\begin{appendix}

  \section{\texorpdfstring{\dzero{}}{D0} cuts}\label{sec:dzero}

  Irrespective of the very clear signal event structure, we employ the
  fast detector simulation \prog{Delphes} as a framework to calculate
  the signal efficiency.  For \dzero we use a simplified calorimeter
  layout composed from cells of dimension $0.1\times\frac{2\pi}{64}$
  in $\eta\times\phi$ space covering $|\eta|\le 4.2$ and
  $\phi\in[0,2\pi[$ where $\eta$ denotes pseudorapidity and $\phi$ the
  azimuthal angle.

  Jets are constructed employing the iterative midpoint algorithm with
  $R=0.5$.
  
  We adopt the cuts of~\cite{Abazov:2010us} by requiring
  \begin{itemize}
  \item at least two isolated photons with $p_T>\gev{25}$ and $|\eta| <
    1.1$,
  \item the azimuthal angle between $\vec\met$ and the hardest jet, if
    existent, is $< 2.5$,
  \item the azimuthal angles between $\vec\met$ and both photons
    are $> 0.2$,
  \item $\met>\gev{35}$.
  \end{itemize}
  Photons must have 95\% of their energy deposited in the
  electromagnetic calorimeter. For a photon to be isolated, the
  calorimetric isolation variable $I$ defined in~\cite{Abazov:2010us}
  must fulfill $I<0.1$ and the scalar sum of transverse momenta of the
  tracks in a distance of $0.05<R<0.4$ from the photon must be smaller
  than \gev{2}. Here is $R=\sqrt{\Delta \phi^2+\Delta\eta^2}$, where
  $\Delta \phi$ is a track's azimuthal distance from the photon and
  $\Delta \eta$ is the corresponding distance in pseudorapidity.

  Note that nearly all signal events fulfill the isolation criteria
  hinting that the hadronic stop decay products are well  separated from
  the photon stemming from the subsequent neutralino decay. This is
  expected as the neutralino decay products $\gamma$ and $\grav$ are
  massless and thus the photon can have a large $p_T$ relative to the
  stop flight direction.

  Also note that we assume that a primary vertex can be reconstructed,
  and that the photon trajectories point to this vertex. The latter
  assumption should be regarded with care when the neutralino lifetime
  is large.

  \section{ATLAS cuts}\label{sec:atlas}

  We use the default \mbox{ATLAS} detector layout implemented in
  \prog{Delphes} and apply the following simplified cuts:
  \begin{itemize}
  \item At least two isolated photons exist with $p_T>\gev{25}$ and
    $|\eta| < 1.81$, but outside the transition region $1.37 < |\eta|
    < 1.52$,
  \item $\met>\gev{125}$
  \end{itemize}
  
  The \mbox{ATLAS} study employs a tight photon selection criterion on
  photon candidates, where the efficiency to identify a true (prompt)
  photon is approximately 85\% in the kinematic region
  considered~\cite{arXiv:1012.4389,atlas-conf-2010-005}. We mimic this
  selection criterion by removing photons from our Monte Carlo sample
  with a probability of 15\%.

  We consider a photon to be isolated if in a cone of width $R=0.2$,
  the scalar $E_T$ sum is less than \gev{5}. Here we sum the $E_T$ of
  the \prog{Delphes} calorimeter cells and exclude the cell the photon
  is mapped to.

  As for the \dzero case, nearly all signal photons fulfill the
  isolation requirement. For larger $\neutr$ masses, the main
  reduction of signal event numbers stems from the 85\% photon
  selection efficiency.

\end{appendix}

%%%%%%%%%%%%%%%%%%%%%%%%%%%%%%%%%%%%%%%%%%%%%%%%%%%%%%%%%%%%%%%%%%%%%% 

\bibliographystyle{unsrt}

\end{document}